\pdfoutput=1

\documentclass[journal,twoside]{IEEEtran}

\usepackage{flushend}  
\usepackage{ifpdf}
\usepackage{cite}
\ifCLASSINFOpdf
  \usepackage[pdftex]{graphicx}
  \DeclareGraphicsExtensions{.pdf,.jpeg,.png}
\else
  \usepackage[dvips]{graphicx}
  \DeclareGraphicsExtensions{.eps}
\fi
\graphicspath{{./figures/}}
\usepackage[cmex10]{amsmath}
\usepackage{amsfonts}  
\usepackage{amssymb}  
\usepackage{amsthm}
\usepackage{array}
\usepackage[caption=false,font=footnotesize]{subfig}
\usepackage{url}

\newcommand{\vect}[1]{\mathbf{#1}}
\newcommand{\matr}[1]{\mathbf{#1}}
\newcommand{\vsym}[1]{\boldsymbol{#1}}    

\newcommand{\set}[1]{\mathcal{#1}}

\newcommand{\F}{\mathbb{F}}
\newcommand{\R}{\mathbb{R}}

\newcommand{\vlambda}{\vsym{\lambda}}

\newcommand{\Eb}{E_\mathrm{b}}

\newcommand{\dv}{d_\mathrm{v}}
\newcommand{\dc}{d_\mathrm{c}}
\newcommand{\ee}{\mathrm{e}}

\theoremstyle{plain}
\newtheorem{thm}{Theorem} 
\newtheorem{lem}[thm]{Lemma}

\newtheorem{cor}[thm]{Corollary}

\theoremstyle{definition}
\newtheorem{exmp}{Example} 
\newtheorem{ass}{Assumption} 

\theoremstyle{remark}
\newtheorem{remk}{Remark} 

\makeatletter
\newlength \figwidth
\if@twocolumn
\else
\fi
\makeatother

\if@twocolumn  
\else
\fi

\usepackage{hyperref}
\hypersetup{
  pdfinfo={
    Title={{LDPC} Code Density Evolution in the Error Floor Region},
    Author={Brian K. Butler},
    Subject={Information Theory, Error Correction Coding},
    Keywords={LDPC codes; Error Correction; Error Floor; Message Passing Decoding; Near-codeword; Trapping Set; Density Evolution; Linear Analysis}
  }
}

\begin{document}

\title{{LDPC} Code Density Evolution in the Error Floor Region}

\author{Brian~K.~Butler,~\IEEEmembership{Senior~Member,~IEEE,}
        Paul~H.~Siegel,~\IEEEmembership{Fellow,~IEEE}
\thanks{This work previously appeared in Section VI of \url{http://arxiv.org/abs/1202.2826v1}, Feb. 2012 and was removed in later revisions.}
\thanks{The first author is an independent consultant and the second author is with the Department of Electrical and Computer Engineering,
University of California, San Diego (UCSD), La Jolla, CA 92093 USA (e-mail: butler@ieee.org, psiegel@ucsd.edu).}
\thanks{This work was supported in part by the Center for Magnetic Recording Research at UCSD and by the National Science Foundation (NSF) under Grant CCF-0829865 and Grant CCF-1116739.}
}

\maketitle

\ifCLASSOPTIONpeerreview
	\markboth{{LDPC} Code Density Evolution in the Error Floor Region}%
	{{LDPC} Code Density Evolution in the Error Floor Region}
\else
	\markboth{DRAFT \;\;\;\;\;\;\;\;Version:  September 19, 2014}%
	{Butler and Siegel: {LDPC} Code Density Evolution in the Error Floor Region}
\fi
%

\begin{abstract} 
This short paper explores density evolution (DE) for low-density parity-check (LDPC) codes at signal-to-noise-ratios (SNRs) that are significantly above the decoding threshold.
The focus is on the additive white Gaussian noise channel and LDPC codes in which the variable nodes have regular degree.

Prior work, using DE, produced results in the error floor region which were asymptotic in the belief-propagation decoder's log-likelihood ratio (LLR) values.
We develop expressions which closely approximate the LLR growth behavior at moderate LLR magnitudes.
We then produce bounds on the mean extrinsic check-node LLR values required, as a function of SNR, such that the growth rate of the LLRs exceeds that of a particular trapping set's internal LLRs such that its error floor contribution may be eliminated.
We find that our predictions for the mean LLRs to be accurate in the error floor region, but the predictions for the LLR variance to be lacking beyond several initial iterations.
\end{abstract}
%

\begin{IEEEkeywords}
Low-density parity-check (LDPC) code, belief propagation (BP), sum-product algorithm (SPA) decoding, 
density evolution, error floor, Margulis code, trapping set.
\end{IEEEkeywords}

%
\IEEEpeerreviewmaketitle

\section{Introduction} 
\IEEEPARstart{L}{ow}-density parity-check (LDPC) codes, first described by Gallager \cite{Gal62}, are an important class of modern error correcting codes.
As a function of channel quality, the error rate performance of an LDPC code under iterative decoding is typically divided into two regions.
The first region, called the \emph{waterfall} region, occurs at poorer channel quality, close to the decoding threshold, and is characterized by a rapid drop in error rate as channel quality improves.
The second region, called the \emph{error floor} region, is the focus of the present paper.
The error floor appears at higher channel quality and is characterized by a more gradual decrease in error rate as channel quality improves.
For message-passing iterative decoders operating on a Tanner graph representing the  code,  the error floor is largely attributed to the effect of small structures, often referred to as trapping sets (TSs), within the Tanner graph.

The understanding of the LDPC error floor in the additive white Gaussian noise (AWGN) channel has progressed significantly, but a number of important challenges, such as the accurate prediction of the error floor for an arbitrary LDPC code, remain.
For Margulis-type codes, MacKay and Postol found that the error floors  observed when transmitting over the additive white Gaussian noise (AWGN) channel  with sum-product algorithm (SPA) decoding were associated with substructures in the Tanner graph that they called \emph{near-codewords}  \cite{MKfloor}.
Shortly thereafter, Richardson wrote a seminal paper on the error floors of memoryless channels \cite{RichFloors}. For a specified decoder operating on a Tanner graph, he used the term \emph{trapping sets} (TSs) to denote the sets of variable nodes responsible for decoding failures in the error floor region.  Both near-codewords and TSs are characterized by a pair of parameters,  $(a, b)$, associated with their corresponding subgraphs, where 
$a$ is the number of variable nodes and $b$ is the number of odd-degree check nodes.
The $(a, b)$ parameters of error-floor-causing structures are typically small.
Several other significant efforts addressing error floors in the AWGN channel exist, including \cite{DolecekTIT,XiaoErrorEstim}.

The present work concerns an aspect of predicting error floors in the AWGN channel.
Sun developed a model of elementary TS behavior based on a linear state-space model of the TS with
density evolution (DE) applied to the code outside of the TS \cite{SunPhD,SunAller}.
An interesting case in Sun's work was the regular-degree, infinite-length LDPC code with a variable-node degree of at least three.
When the SPA decoder's metrics were allowed to grow very large, Sun showed that the Tanner graph outside of the TS 
will eventually correct an elementary TS received in error.
Schlegel and Zhang \cite{SCH10} extended Sun's work by adding time-variant gains
and channel errors from outside of the TS to the model, while log-likelihood ratio (LLR) message values were permitted to vary within a limited range.
Butler and Siegel \cite{ButlerAller,ButlerFloor} further refined the linear state-space model and explored several cases in which the error floors are reduced by many orders of magnitude by reducing the extent of saturation (or ``clipping'') highly certain messages within the SPA decoder, including the $(2640,1320)$ Margulis code.
Zhang and Schlegel \cite{SCH13} have independently proposed refinements and similarly examined major error floors reductions.

Sun's analytical development was asymptotic in LLR values in addition to the usual asymptotic assumptions from DE on the block length and number of iterations.
Herein, we work to identify the conditions when the growth of the check nodes' beneficial messages from outside of the TS 
can out-pace the internal growth of the TS's detrimental messages.
In doing so, we develop new signal-to-noise ratio (SNR) thresholds that indicate when this mean LLR growth regime is reached.
Unfortunately, the model of error floor prediction also depends upon the variance of the LLRs, and it is in this respect that our DE analysis is inconclusive.


\section{Preliminaries} 
\label{sect-pre}
\label{ss-awgnspa}
The column vector $\vect{c}$ is a codeword of the LDPC code $\set{C}$ if $\vect{c}$ satisfies $\matr{H} \vect{c} = \vect{0}$,
where $\matr{H}$ is a sparse parity-check matrix whose entries are elements of a particular field.

\begin{ass}
We are only concerned with LDPC codes over the binary field $\F_2$. 
We consider only codes described by $\dv$-variable-regular Tanner graphs with $\dv \ge 3$.
\end{ass}

\begin{ass}
We assume binary antipodal signaling over the AWGN channel, which
is characterized by its SNR $1/\sigma^2$ or $\Eb / N_0$,
in which $1/\sigma^2 = 2 R \Eb / N_0$, where $R$ is the rate of the code in information bits per binary symbol.
\end{ass}

\begin{ass}
The TSs to be studied are elementary.
\end{ass}

A TS is called \emph{elementary} if all the check nodes in its induced Tanner subgraph are of degree one or two \cite{Laendner09}.
Elementary TSs are simple enough to be modeled with linear systems
and also account for majority of TSs contributing to the error floors of belief propagation decoders seen in practice 
\cite{Han09,RichFloors,Laendner09,Cole,ZhangRyanFloor,Milenk07}.

First, the state vector $\vect{x}_l$ of message values in the LLR domain is initialized using the vector of intrinsic information from the channel $\vlambda$ according to $\vect{x}_0 = \matr{B} \vlambda$.
Then, $\vect{x}_l$ is updated once per full iteration of SPA decoding in the LLR-domain according to
\begin{equation}
\label{SS2}
\vect{x}_l = \bar{g}_l' \matr{A} \vect{x}_{l-1} + \matr{B} \vlambda + \matr{B}_{\mathrm{ex}} \vlambda_l^{(\mathrm{ex})}
\end{equation}
for iterations $l\ge1$, where $\vlambda_l^{(\mathrm{ex})}$ is the vector of LLR messages extrinsic to the TS \cite{SunPhD,SunAller,SCH10,SCH13,ButlerAller,ButlerFloor}. 
The central part of the state-space model is the updating of the state vector $\vect{x}_l$.
The elements of $\vect{x}_l$ are the LLR messages sent along the edges from the subgraph's variable nodes toward the degree-two check nodes.
The message processing of the degree-two check nodes in the subgraph are modeled with a mean modified gain $\bar{g}_l'$, which is time varying and approaching $1$ as the LLR magnitudes increase outside the TS \cite{ButlerAller,ButlerFloor}.
The vector $\bar{g}_l' \vect{x}_{l-1}$ represents the LLR messages produced by the degree-two check nodes 
during the first half of iteration $l$,  and $\bar{g}_l' \matr{A} \vect{x}_{l-1}$ represents their contribution to the variable-node update during the second half of iteration $l$.
Finally, based on the structure of the subgraph induced by the TS under study set we may construct the $(0,1)$-matrices $\matr{A}$, $\matr{B}$, and $\matr{B}_{\mathrm{ex}}$ as described in \cite{ButlerFloor}.

In \cite{SCH13,ButlerFloor}, the linear model of \eqref{SS2} accurately predicted the likelihood with which the TSs fail when
the decoder saturates LLRs for several codes, but here we hope to gain insight into the case of the non-saturating decoder.
The dominant eigenvalue of $\matr{A}$, denoted by $r=\rho(\matr{A})$, is useful to bound the growth rate of $\vect{x}_l$ in \eqref{SS2}.
For $\dv$-variable-regular codes, it is easy to show that $r < \dv-1$ \cite{SunPhD,ButlerFloor}.
One may apply DE to model the contributions from the nodes outside of the TS denoted by $\vlambda_l^{(\mathrm{ex})}$ in \eqref{SS2}.
In SPA decoding this contribution appears as the LLR messages sent from the degree-one (unsatisfied) check nodes toward their neighboring variable nodes.

\section{Analysis Using DE without Saturation} 
\label{sect-analnosat}

DE was introduced in \cite{RUCAP,RUDE} as a technique to estimate the distribution of messages in an SPA decoder
and to find the {decoding threshold} of ensembles of LDPC codes with specific degree distributions in the limit 
as the block length and the number of iterations tend to infinity.
For large block lengths, the error rate typically drops very dramatically as the SNR exceeds the decoding threshold.
As we are interested in the behavior of the error floor in this work, we assume that the channel SNR is always above the decoding threshold.

In the limit, as assumed in DE, the block length goes to infinity and the Tanner graph is free of cycles.
However, in any ``good" finite-length code, the Tanner graph will indeed have cycles \cite{cyclefree}.
On the other hand, we generally find the SPA decoder does quite well even though it implicitly uses the cycle-free assumption.


Applying DE with SNRs greater than the decoding threshold, Sun showed,
in \cite{SunPhD}, that mean check-node output LLR, $m_{\lambda(\mathrm{ex})}$, grows from iteration $l-1$ to $l$ as
\begin{equation*}
m_{\lambda(\mathrm{ex})}^{(l)}=(\dv -1) m_{\lambda(\mathrm{ex})}^{(l-1)} + \mbox{``some~small~value~terms."}
\end{equation*}
We seek to bound the effect of Sun's ``small value terms," by developing the governing expressions rather than the asymptotic result.
We will find bounds on the SNR region in which we can expect the growth in $m_{\lambda(\mathrm{ex})}^{(l)}$ necessary to eliminate error floors.

We begin by using the consistent Gaussian approximation to DE for the AWGN channel \cite{RUDE}.
The update rule at iteration $l$ for the check-nodes' output mean $m_{\lambda(\mathrm{ex})}$ is 
\begin{equation}
\label{DE1}
m_{\lambda(\mathrm{ex})}^{(l)}=\phi^{-1}\left(1-\left[1-\phi\left(m_{\lambda} + (\dv -1)m_{\lambda(\mathrm{ex})}^{(l-1)}\right)\right]^{\dc -1}\right),
\end{equation}
where $\phi (x)$ is defined by the integral
\begin{equation*}
\phi (x) \triangleq \begin{cases}
1-\frac{1}{\sqrt{4\pi x}} \int_{\R} \tanh \frac{u}{2}~\ee^{-\frac{(u-x)^2}{4x}}du , &\mbox{if $x>0$}\\
1, &\mbox{if $x=0$}.
\end{cases}
\end{equation*}
Note that $\phi (x)$ is continuous and monotonically decreasing from $1$ to $0$, as $x$ increases from $0$ toward $\infty$.

For any $x>0$, the function $\phi(x)$ is bounded \cite{RUDE} by 
\begin{equation}
\label{DE_PB0}
\sqrt{\frac{\pi}{x}} \ee^{-x/4} \left(1-\frac{3}{x}\right) < \phi(x) < \sqrt{\frac{\pi}{x}} \ee^{-x/4} \left(1-\frac{1}{7x}\right).
\end{equation}
These bounds tighten significantly as $x$ increases.
We will take the lower bound of (\ref{DE_PB0}) and use it to upper bound $1- \phi(x)$ as
\begin{equation}
\label{DE_PB2}
1-\delta \sqrt{\frac{\pi}{x}} \ee^{-x/4} > 1-\phi(x),
\end{equation}
where $\delta \triangleq 1-3/x$, noting that $\delta$ approaches $1$ for large $x$.
We take a small digression before proceeding with the main theorems to develop inequalities that we will require.

\begin{lem}
\label{L0}
For any $\alpha \in \R$ and $x,\beta> 0$ there exists a unique $y \in \R$ such that $\alpha=y+\beta\ln \left(1+y/x\right)$.
Moreover, $y>-x$.
\end{lem}
\begin{IEEEproof}
The function $f(y)=y+\beta\ln \left(1+y/x\right)$ is well-defined only on the domain $y>-x$.
It is continuous and monotonically increasing on that domain, and $f(y) \rightarrow -\infty$ as $y \rightarrow -x$ from the right.
\end{IEEEproof}

\begin{lem}
\label{L1}
Let $m+\beta \ln m > x + \beta \ln x + \alpha$ where $m,x,\alpha,\beta \in \R$ with $m,x,\beta> 0$.
Then $m > x + \frac{\alpha}{1+\beta/x}$.
\end{lem}
\begin{IEEEproof}
First we will prove by contrapositive that $m > x + y$, where $\alpha=y+\beta\ln \left(1+y/x\right)$. 
From Lemma~\ref{L0} we know for any $\alpha$ we can find a unique real $y > -x$.

Assume that $m \le x+ y$. Then we can show $\beta \ln m \le \beta \ln \left( x+ y \right) = \beta \ln x+ \beta \ln \left( 1+ y/x \right)$.
Adding these two inequalities term-by-term yields
\begin{equation*}
m + \beta \ln m \le x + \beta \ln x+ y + \beta \ln \left( 1+ y/x \right).
\end{equation*}
Recognizing that the right two terms combine to equal $\alpha$, we have an expression that exactly contradicts our original proposition.
Thus, $m > x+ y$.

Since we know for any $z>-1$ that $z \ge \ln(1+z)$, we can upper bound $\alpha$ as
\begin{equation*}
\alpha=y + \beta \ln \left( 1+y/x \right) \le y + \beta y/x. 
\end{equation*}
Solving for $y$ yields $y \ge \frac{\alpha}{1+\beta /x}$.
Finally, we combine these two inequalities to produce the desired result, \textit{i.e.},
\begin{equation*}
m > x+ y \ge x + \frac{\alpha}{1+\beta /x}.
\end{equation*}
\end{IEEEproof}
\begin{remk}
The bound of Lemma~\ref{L1} is tightest when $m$ and $x$ are of similar magnitude, \textit{e.g.}, $0.5<m/x<2$.
\end{remk}

\begin{thm}
\label{TDE}
Given a $(\dv,\dc)$-regular LDPC code, with $\dv \ge 3$, at sufficiently high $\Eb/N_0$, after a finite number of iterations, DE predicts that the SPA decoder's check-node output mean $m_{\lambda(\mathrm{ex})}^{(l)}$ dominates $(\dv -1)^l$ asymptotically.
\end{thm}

\begin{IEEEproof}
First we will assume that the ${\Eb}/{N_0}$ exceeds the decoding threshold, so we are assured that the LLRs will continue growing by DE.
We start by restating (\ref{DE1}) in the form
\begin{equation}
\label{DE2}
1-\phi \left(m_{\lambda(\mathrm{ex})}^{(l)}\right)=\left[1-\phi\left(m_{\lambda} + (\dv -1)m_{\lambda(\mathrm{ex})}^{(l-1)}\right)\right]^{\dc -1}.
\end{equation}
Next, by applying the upper bound of (\ref{DE_PB2}) to the left-hand side (LHS) of (\ref{DE2}), where $x=m_{\lambda(\mathrm{ex})}^{(l)}>0$, we have
\begin{equation*}
\begin{split}
1&-\delta \sqrt{\frac{\pi}{ m_{\lambda(\mathrm{ex})}^{(l)}}}~ \ee^{- m_{\lambda(\mathrm{ex})}^{(l)}/4} > \\
&\left[1-\phi\left(m_{\lambda} + (\dv -1)m_{\lambda(\mathrm{ex})}^{(l-1)}\right)\right]^{\dc -1},
\end{split}
\end{equation*}
where $\delta = 1 - 3/m_{\lambda(\mathrm{ex})}^{(l)}$.

Now we can truncate the binomial expansion to a simple bound, in the form shown below for small $\phi(x)$.
Specifically, for any positive integer $n$,
\begin{equation}
\label{binom}
\begin{split}
\left(1-a\right)^{n}&= 1-na+ \binom{n}{2} a^2 - \binom{n}{3} a^3 \ldots+ (-a)^n \\
& > 1-na, \quad \mbox{for small enough $a$}.
\end{split}
\end{equation}
For example, by the time the LLRs have reached $m_{\lambda} + (\dv-1) m_{\lambda(\mathrm{ex})}^{(i-1)}>8$, we are assured that 
$\phi(\cdot) < 0.09$ by (\ref{DE_PB0}), which is small enough for the truncated binomial (\ref{binom}) to hold as a positive lower bound for $\dc \le 12$.
Codes with higher check degree will require larger LLRs for (\ref{binom}) to hold.
Utilizing this binomial bound we now have
\begin{equation}
\label{DE5}
\sqrt{\frac{\pi}{ m_{\lambda(\mathrm{ex})}^{(l)}}}~ \ee^{- m_{\lambda(\mathrm{ex})}^{(l)}/4} < \frac{\dc-1}{\delta} \phi\left(m_{\lambda} + (\dv -1)m_{\lambda(\mathrm{ex})}^{(l-1)}\right).
\end{equation}

Next, we apply the upper bound of (\ref{DE_PB0}) with $x=m_{\lambda} + (\dv -1)m_{\lambda(\mathrm{ex})}^{(l-1)}$ to the RHS of (\ref{DE5}),
take the logarithm of both sides, and scale by $-4$ to find
\begin{equation}
\begin{split}
\label{DE7}
2& \ln m_{\lambda(\mathrm{ex})}^{(l)} + m_{\lambda(\mathrm{ex})}^{(l)} > \\
&-4 \ln \left( \frac{(\dc-1) \left(1-\frac{1}{7x}\right)}{\delta} \right) + 2\ln x + x.
\end{split}
\end{equation}
Applying Lemma~\ref{L1} with $\beta=2$ to (\ref{DE7}) yields
\begin{equation}
\begin{split}
\label{DE8}
m_{\lambda(\mathrm{ex})}^{(l)} 
&> x - 4\frac { \ln \left\{ \frac{\dc-1}{\delta} \left(1-\frac{1}{7x}\right)\right\} }{1+ 2/x}\\
&= m_{\lambda} + (\dv -1)m_{\lambda(\mathrm{ex})}^{(l-1)} - 4\frac { \ln \left\{ \frac{\dc-1}{\delta} \left(1-\frac{1}{7x}\right)\right\} }{1+ 2/x}.
\end{split}
\end{equation}
Examining (\ref{DE8}), we see that if the $m_{\lambda}$ term exceeds the magnitude of the final term which is negative,
we can guarantee $m_{\lambda(\mathrm{ex})}^{(l)}$ growth by at least a factor of $\dv-1$ at each iteration.
Noting that the mean channel LLR is $m_{\lambda} = 4 R \Eb / N_0$ in the AWGN channel, 
the condition needed to satisfy $m_{\lambda(\mathrm{ex})}^{(l)} > C (\dv -1)^l$ is
\begin{equation}
\label{DE9}
R \frac{\Eb}{N_0} > \frac { \ln \left\{ \frac{\dc-1}{\delta} \left(1-\frac{1}{7x}\right)\right\} }{1+ 2/x},
\end{equation}
where $x=m_{\lambda} + (\dv -1)m_{\lambda(\mathrm{ex})}^{(l-1)} \gg 0$, and $x$ is increasing with each iteration.
Since the two factors containing $x$ reduce the RHS of (\ref{DE9}) for all $x>1/7$, we can make the condition (\ref{DE9}) simpler, but stronger as
\begin{equation}
\label{DE10}
R \frac{\Eb}{N_0} > \ln \frac {\dc-1}{\delta},
\end{equation}
where $\delta = 1 - 3/ m_{\lambda(\mathrm{ex})}^{(l)} > 0$.
The value of $\delta$ approaches $1$ as the number of iterations increases.
When the decoder reaches an iteration such that our earlier assumptions on the magnitude of the LLRs are satisfied and if \eqref{DE10} is satisfied, 
then LLR growth will satisfy $m_{\lambda(\mathrm{ex})}^{(l)} > (\dv -1) m_{\lambda(\mathrm{ex})}^{(l-1)}$ for all future iterations.
\end{IEEEproof}

With $\delta=1$, \eqref{DE10} has appeared in \cite[p.~38]{Wiberg}, \cite{LentBlockErr,KoetVont} for several LDPC decoder algorithms as a breakout channel condition in which the error rate tends toward zero as the iteration count and block length increase toward infinity.

\begin{exmp}
For any $(3,6)$-regular code, $R=0.5$, we compute the ${\Eb}/{N_0}$ threshold of (\ref{DE10}) to be $5.077$ dB, assuming that $\delta\approx1$.
At any ${\Eb}/{N_0}$ greater than $5.077$ dB, we can guarantee that $m^{(i)}_{\lambda(\mathrm{ex})} > C (\dv-1)^i$ is reached after enough iterations
to satisfy the assumptions made in this section.
This is a sufficient extrinsic LLR growth rate to overcome any potential elementary TS, because the TS's LLR growth rate is limited by $r < \dv-1$.  
However, we have only guaranteed it in a mean sense.
\end{exmp}

\begin{cor}
\label{cor1}
Given a $(\dv,\dc)$-regular LDPC code, with $\dv \ge 3$, at any $\Eb/N_0$ above the decoding threshold, after a finite number of iterations $l$, DE predicts that the SPA decoder's check-node output mean $m_{\lambda(\mathrm{ex})}^{(l)}$ dominates $r^l$ asymptotically, for any $r$, such that $0 \le r < \dv-1$.
\end{cor}
\begin{IEEEproof}
The condition we want to prove is
\begin{equation}
\label{DE11}
\frac{m_{\lambda(\mathrm{ex})}^{(l)}} {m_{\lambda(\mathrm{ex})}^{(l-1)}} > r,
\end{equation}
which, by (\ref{DE8}) of the previous proof, will be true if the following stronger condition holds
\begin{equation}
\label{DE12}
\dv-1 + \frac{m_{\lambda}} {m_{\lambda(\mathrm{ex})}^{(l-1)}} - \frac{4} {m_{\lambda(\mathrm{ex})}^{(l-1)}} 
\frac { \ln \left\{ \frac{\dc-1}{\delta} \left(1-\frac{1}{7x}\right)\right\} }{1+ 2/x} > r,
\end{equation}
where $ x=m_{\lambda} + (\dv -1)m_{\lambda(\mathrm{ex})}^{(l-1)}$.
Define $\epsilon$ such that $\epsilon \triangleq \dv-1-r>0$.
Rearranging terms, we have the equivalent condition
\begin{equation}
\label{DE13}
R \frac{\Eb}{N_0} > \frac { \ln \left\{ \frac{\dc-1}{\delta} \left(1-\frac{1}{7x}\right)\right\} }{1+ 2/x} - \frac{\epsilon \; m_{\lambda(\mathrm{ex})}^{(l-1)}}{4},
\end{equation}
which will be true at some iteration $l$ given any positive growth in $m_{\lambda(\mathrm{ex})}^{(l)}$.
\end{IEEEproof}

\begin{figure}
\centering
\includegraphics[width=\figwidth]{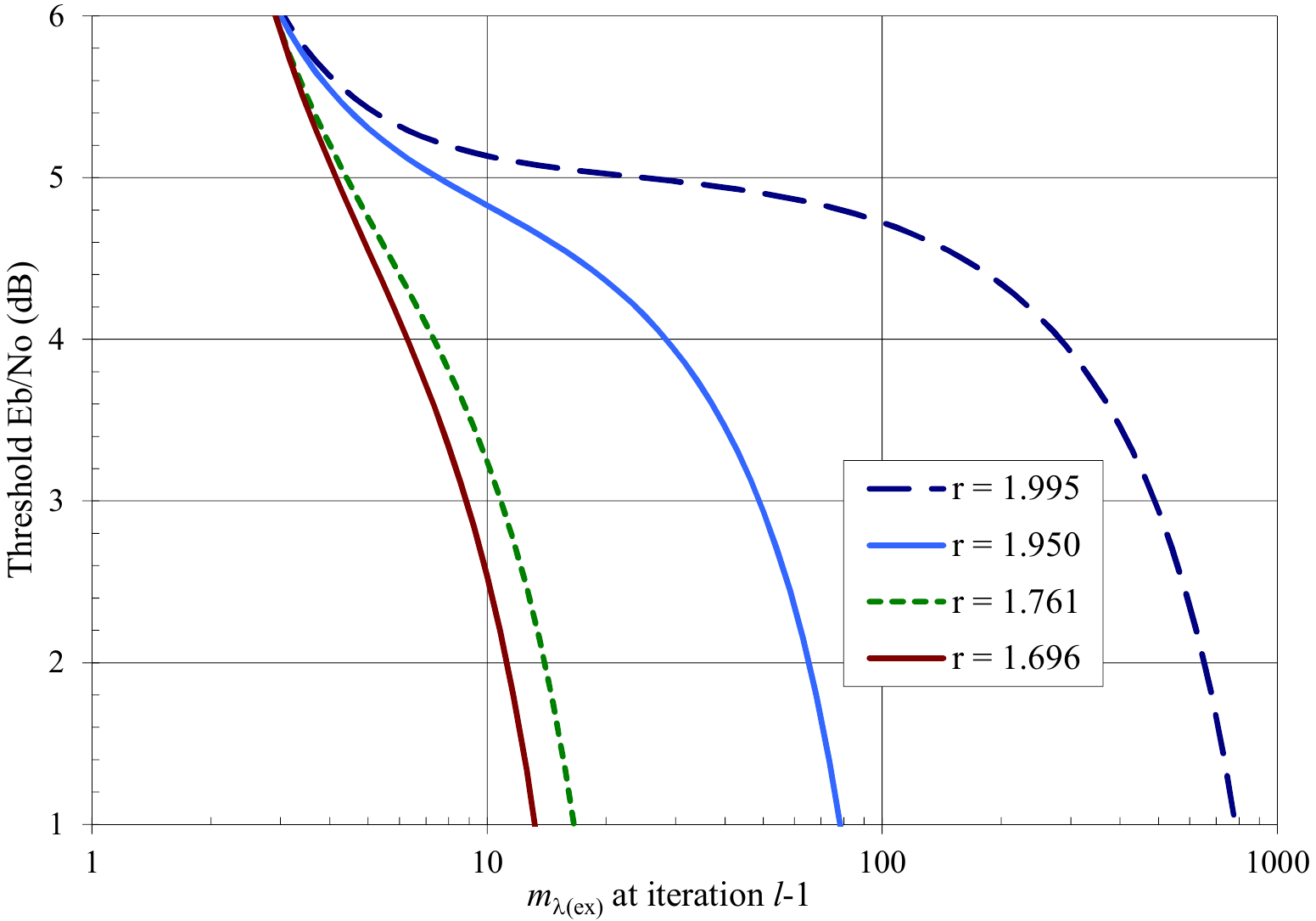}
\caption{Threshold $\Eb/N_0$ in dB vs. $m_{\lambda(\mathrm{ex})}^{(l-1)}$ to achieve the LLR growth needed to overcome
a TS characterized by $r$ as specified in the legend for $(3, 6)$-regular codes such as Margulis.}
\label{fig_EbThresh}
\end{figure}

Once received erroneously, the most difficult elementary TSs to overcome, that are not codewords, have a spectral radius very close to $\dv-1$, which is the upper bound on spectral radii \cite{ButlerFloor}.
To approach such an extreme spectral radius, the TS would require a very large $a$ and small $b$ parameter.
The ${\Eb}/{N_0}$ required to reach the LLR growth regime of $m^{(i)}_{\lambda(\mathrm{ex})} > C r^i$, which is necessary to overcome the trapping 
behavior of a specific TS, may be significantly less than the threshold of (\ref{DE10}) as the next example illustrates.

\begin{exmp}
Continuing the prior example, we note that the $(2640, 1320)$ Margulis code is $(3,6)$-regular, with the most troublesome TSs
reportedly being $(12,4)$ and $(14,4)$ \cite{Han09,RichFloors,MKfloor}.
We have computed their maximum eigenvalues to be $r\approx 1.696$ and $r\approx 1.761$, respectively.
Fig.~\ref{fig_EbThresh} shows the behavior of the threshold pair $\left(\Eb/N_0,m_{\lambda(\mathrm{ex})}^{(l-1)}\right)$, that is required by (\ref{DE13}) to
achieve a growth rate in the mean equivalent to the specific eigenvalues mentioned and two hypothetical extreme eigenvalues of $1.95$ and $1.995$.
Note that this figure suffers some approximation errors at the smallest LLRs shown.
From Fig.~\ref{fig_EbThresh} we see that for the $(12,4)$ TS at an $\Eb/N_0$ of $2.8$ dB we will require $m_{\lambda(\mathrm{ex})}^{(l-1)} > 9.3$
to enter the LLR growth regime of $m^{(i)}_{\lambda(\mathrm{ex})} > C (1.696)^i$ necessary to eventually overcome that TS.
Using simulation or DE, one can find that this mean LLR condition is satisfied at the start of the eighth iteration.
\end{exmp}

\begin{thm}
\label{TDEi}
Given a variable-regular, but check-irregular ($\dv,\rho(x)$) LDPC code, with $\dv \ge 3$, at sufficiently high $\Eb/N_0$,
after a finite number of iterations, DE predicts that the SPA decoder's check-node output mean $m_{\lambda(\mathrm{ex})}^{(l)}$ 
dominates $ (\dv -1)^l$ asymptotically.
\end{thm}
\begin{IEEEproof}
Let $d_{\mathrm{r}}$ denote the maximum check degree for the graph and let $\rho(x)$ represent a polynomial with coefficients
$\rho_j$, $j\in\{2,\ldots ,d_{\mathrm{r}}\}$, denoting the fraction of edges belonging to degree-$j$ check nodes.
If we separate $m_{\lambda(\mathrm{ex})}^{(l)}$ into its components we can reuse our results from Theorem~\ref{TDE}.
Given
\begin{equation}
\label{DEi1}
m_{\lambda(\mathrm{ex})}^{(l)} =\sum^{d_{\mathrm{r}}}_{j=2} \rho_j m_{j,\lambda(\mathrm{ex})}^{(l)}, 
\end{equation}
where
\begin{alignat}{2}
m_{j,\lambda(\mathrm{ex})}^{(l)} &= \phi^{-1}\left(1-\left[1-\phi\left(x\right)\right]^{j-1}\right),\\
x &= m_{\lambda} + (\dv -1)m_{\lambda(\mathrm{ex})}^{(l-1)},
\end{alignat}
and letting $\delta_j \triangleq 1-3/m_{j,\lambda(\mathrm{ex})}^{(l)}$, (\ref{DE8}) implies
\begin{equation}
\begin{split}
\label{DEi2}
m_{\lambda(\mathrm{ex})}^{(l)} 
&> m_{\lambda} + (\dv -1)m_{\lambda(\mathrm{ex})}^{(l-1)} \\
&\;\; - \sum^{d_{\mathrm{r}}}_{j=2} \rho_j \frac {4 \ln \left\{ \frac{j-1}{\delta_j} \left(1-\frac{1}{7x}\right)\right\} }{1+ 2/x}.
\end{split}
\end{equation}
As before, the growth condition is always met with an even stronger condition on the SNR, which is
\begin{equation}
\label{DEi3}
R \frac{\Eb}{N_0} > \sum^{d_{\mathrm{r}}}_{j=2} \rho_j \ln \frac{j-1}{\delta_j}.
\end{equation}
When (\ref{DEi3}) is satisfied, then $m_{\lambda(\mathrm{ex})}^{(l)} > (\dv -1) m_{\lambda(\mathrm{ex})}^{(l-1)}$.
\end{IEEEproof}

\begin{cor}
Given a variable-regular, but check-irregular ($\dv,\rho(x)$) LDPC code,
with $\dv \ge 3$, at any $\Eb/N_0$ above the decoding threshold, after a finite number of iterations $l$,
DE predicts that the SPA decoder's check-node output mean $m_{\lambda(\mathrm{ex})}^{(l)}$ dominates
$r^l$ asymptotically, for any $r$, such that $0 \le r < \dv-1$.
\end{cor}
\begin{IEEEproof}
Similar to Corollary~\ref{cor1}.
\end{IEEEproof}

When combined with the model developed in \cite{ButlerFloor}, the results of this section suggest that every potential (non-codeword) elementary TS error can be corrected by a non-saturating LLR-domain SPA decoder after a sufficient number of iterations, provided certain assumptions hold \cite{ButlerFloor}.
These assumptions include that the distribution of the extrinsic LLRs is approximately Gaussian and the grow rate of the LLR variance $\sigma_{l(\mathrm{ex})}^{2}$ is no greater than the growth rate of $m_{\lambda(\mathrm{ex})}^{(l)}$.
These assumptions fit those used in justifying the consistent Gaussian approximation to AWGN DE\cite{RUDE}, which assumes that the Tanner graph does not contain cycles since it takes the statistical distributions to be independent among incoming edges.
Because useful finite-length LDPC codes \emph{do} contain cycles, it is worth questioning the use of DE to predict error floors.

\begin{figure}
\centering
\includegraphics[width=\figwidth]{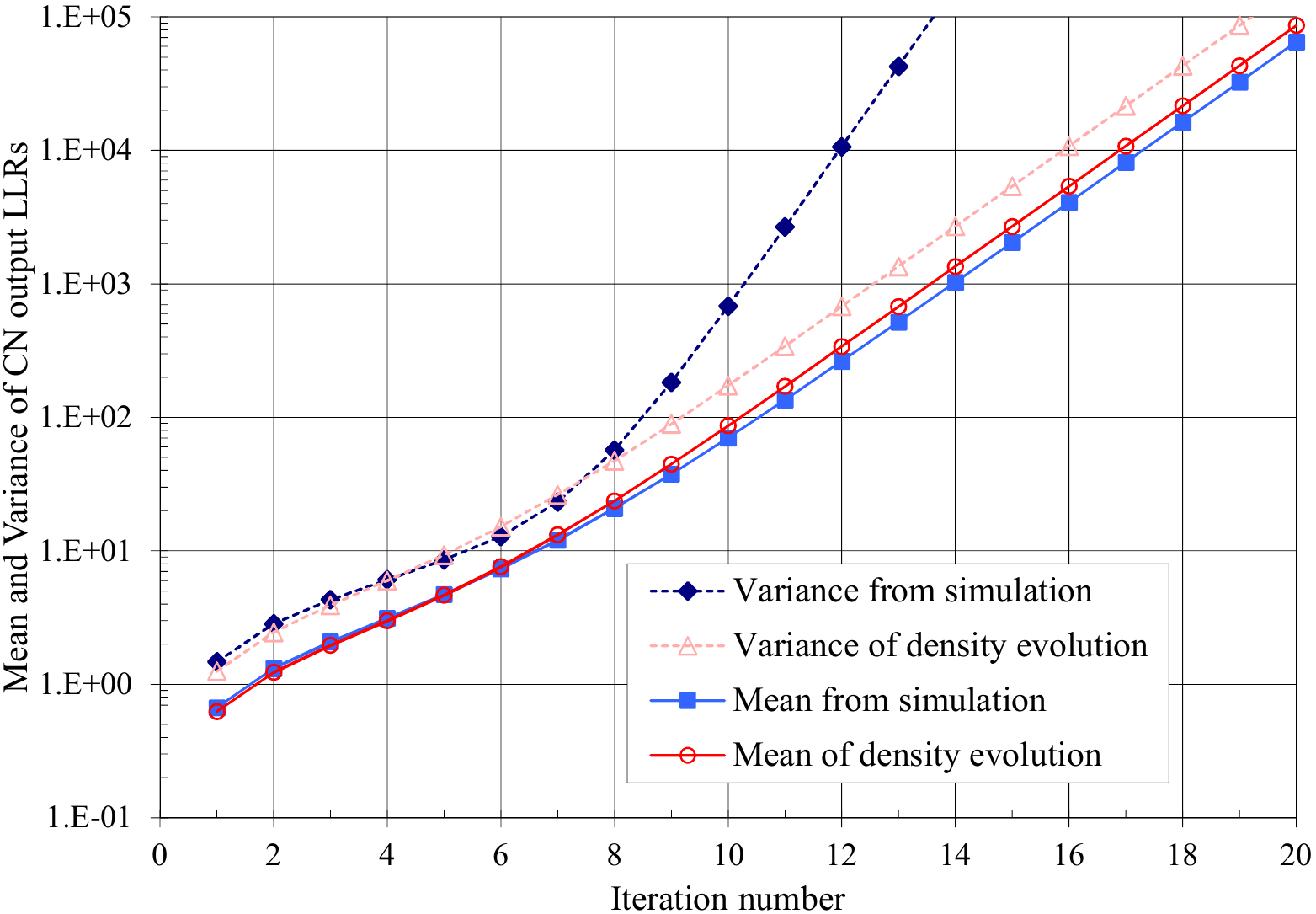}
\caption{Check-node output LLR mean and variance versus iteration number for the Margulis code at $\Eb/N_0$ of $2.8$ dB.}
\label{fig_LLRG}
\end{figure}

In Fig.~\ref{fig_LLRG} we plot the mean and variance of check-node output LLRs from the SPA decoder simulation of the $(2640, 1320)$ Margulis code
and from DE of a $(3,6)$-regular code.
In the case of the simulation, the all-zero codeword is transmitted over the AWGN channel and early termination
of the LLR-domain SPA decoder is disabled.
We note that the mean follows very closely the mean LLR predicted by DE.
This mean also follows very closely the lower bounds we developed in this section, once the LLR is greater than ten.
The variance of the check-node output LLRs, however, shows a problem. 
For the first seven iterations, the variance is approximately twice the mean, as predicted by the Gaussian approximation to DE.
However, by the ninth iteration, the variance has clearly taken on the trend of the square of the mean.
This trend can generate an error floor in the model as the mean to standard deviation ratio of the error indicator, denoted by $\beta'_l$ in \cite{ButlerAller,ButlerFloor}, reaches a fixed value rather than growing as iterations get large.

\section{Conclusion} 

With respect to Sun's work in \cite{SunPhD}, we are in agreement with Sun's conclusion as applied to variable-regular LDPC codes without cycles.
While Sun's analysis was asymptotic, ours used bounds to derive the SNR thresholds and SNR-LLR regions in which the beneficial 
metrics from the TS's degree-one check nodes grow fast enough to overcome the faulty channel information that triggers a TS failure.
Further, we have shown empirically for codes with cycles that without saturation the ratio of the mean to standard deviation converges 
to a finite value, which may imply an error floor.

We are hopeful that additional efforts to accurately model the evolution of the LLR distribution (in the presence of cycles) will yield further insight into error floor behavior.

%



\bibliographystyle{IEEEtran}
\bibliography{IEEEabrv,Butler}


\end{document}